\documentstyle[preprint,revtex]{aps}
\begin{document}
\draft

\begin{title}
A numerical transfer-matrix study of \\
surface-tension anisotropy in Ising models on \\
square and cubic lattices
\end{title}
\author{
        Howard~L.~Richards*\ddag,
        M.~A.~Novotny\dag,
        and Per~Arne~Rikvold*\dag\ddag \\~}
\begin{instit}
      *Department of Physics B-159,\\
      \dag Supercomputer Computations Research Institute B-186,\\
      and \ddag Center for Materials Research and Technology B-159,\\
      Florida State University, \\
      Tallahassee, FL 32306, U.S.A.
\end{instit}
%\receipt{May 7,1993}

%%%%%%%%%%%%%%%%%%%%%%%%   ABSTRACT   %%%%%%%%%%%%%%%%%%%%%%%%

\begin{abstract}

We compute by numerical transfer-matrix
methods the surface
free energy $\tau(T),$ the surface stiffness
coefficient $\kappa(T),$ and the single-step free
energy $s(T)$ for
Ising ferromagnets with $(\infty \! \times \! L)$ square-lattice
and $(\infty \! \times \! L \! \times \! M)$ cubic-lattice
geometries, into which an interface
is introduced by imposing antiperiodic or plus/minus
boundary conditions in one transverse direction.
The surface tension $\sigma(\theta,T)$
per unit length of interface projected  % with angle $\theta$
in the longitudinal direction defines $\tau(T)$ and
$\kappa(T)$ for temperatures at which the interface
is rough:
\( \sigma(\theta,T)/ \! \cos\theta
  = \tau(T) + {1\over 2}\kappa(T)\theta^2 +
  O(\theta^4).
\)
For temperatures at which the interface is smooth,
$s(T)$ is the free-energy contribution of the
dominant entropy-producing fluctuation, a single-step
terrace.
The finite-size scaling behavior of the interfacial
correlation length provides the means of investigating
$\kappa(T)$ and $s(T).$
The resulting transfer-matrix estimates are fully consistent
with previous series and Monte Carlo studies, although
current computational technology does not permit
transfer-matrix studies of sufficiently large systems
to show quantitative improvement over the previous estimates.
\end{abstract}

%%%%%%%%%%%%%%%%%%%%%%%% END ABSTRACT %%%%%%%%%%%%%%%%%%%%%%%%

\pacs{68.10.Cr, 68.35.Md, 05.50.+q, 11.15.Ha}
% 68.10.Cr   : Surface energy (surface tension, interface
%              tension, angle of contact, etc.) (FLUIDS)
% 68.35.Md   : Surface energy; thermodynamic properties (SOLIDS)
% 05.50.+q   : Lattice theory and statistics; Ising problems
% 11.15.Ha   : Lattice gauge theory
\narrowtext
%%%%%%%%%%%%%%%%%%%%%%%%   SECTION 1  %%%%%%%%%%%%%%%%%%%%%%%%
%%%%%%%%%%%%%%%%%%%%%%%% Introduction %%%%%%%%%%%%%%%%%%%%%%%%

\section{Introduction}
  \label{sec-Intro}

Many phenomena of interest in condensed-matter physics,
materials science, and high-energy physics occur in systems
in which two or more thermodynamic phases coexist
below some critical temperature $T_c.$ These phenomena
are strongly influenced by the behavior of interfaces
between the phases.
Examples of interfacial phenomena from condensed-matter physics include
wetting, nucleation and growth, and crystal faceting.
An example from high-energy physics is spatially non-uniform
symmetry-breaking in the early universe.
A number of these systems also have a {\em roughening transition}
at a temperature $T_R$ (known as the roughening temperature),
at which the interface changes from
microscopically flat to microscopically rugged
\cite{RoughRev,Chui}.
For the three-dimensional (3D) cubic-lattice Ising model,
recent Monte Carlo studies yield
\(T_R \! \approx \! (0.542 \pm 0.005)T_c \approx 2.45 J/k_B\)
\cite{HasenPhD,MRS5,MCRGTc}.
The roughening transition of one exactly solvable
Solid-on-Solid (SOS) model \cite{Beij77}
has been shown to belong to the
Kosterlitz-Thouless (KT) universality class \cite{KT73,KT74},
and roughening transitions in related SOS models
and in the 3D Ising model are generally believed to
belong to this universality class as well \cite{RoughRev}.
On the other hand, for the two-dimensional (2D)
square-lattice Ising model \( T_R \! =  \! 0 \),
that is, the interface is always rough.

An aspect of interfaces that is of great importance in
determining interface profiles and dynamics
is the surface tension, which is the amount by which
each unit area of interface increases the total free energy of the
system in the thermodynamic limit \cite{Gelf88}.
A great deal of analytical work \cite{MRS1a,MRS1b,MRS1c,Kleban91},
including low-temperature series expansions
\cite{Adler87,Holzer89,MRS2},
has been done to study interfaces
and their associated surface tensions.
The surface tensions in a $\phi^4$ field theory \cite{MRS7}
and in the $q$-state Potts model \cite{MRS8,Borgs92}
have been studied by Monte Carlo simulations.
Of particular interest has been the dependence of the surface
tension in the 2D and 3D Ising models on the orientation of
the interface relative to the underlying lattice
(see Fig.~\ref{fig:3Dsys});
this has been
studied both analytically \cite{Priv88,MRS10,MRS11} and
through Monte Carlo simulations
\cite{MRS5,MRS7,Mon88,MRS4,MRS6,CGV92,Pinn92,Berg93}.

The analytical treatments of the 3D Ising model rely on
approximations which are perturbative in nature, whereas
the non-perturbative Monte Carlo method suffers from critical
slowing-down and the possibility of becoming trapped in
metastable or long-lived unstable states.
Furthermore, free energies and correlation lengths are
cumbersome to calculate in Monte Carlo simulations.
An alternative method which has been successful in the
study of phase transitions, particularly in 2D systems,
is that of numerical transfer-matrix (TM) calculations
combined with finite-size-scaling (FSS) analysis \cite{MRS20}.
The numerical TM FSS method does not encounter the
aforementioned difficulties:
(i)   It is nonperturbative and does not require
      foreknowledge of special temperatures
      (such as $T_R$ and $T_c$) or of critical exponents;
(ii)  It has no dynamics, so that critical slowing-down
      does not occur and the equilibrium state is readily found;
(iii) The free energy corresponds to the largest
      eigenvalue of the TM, and so is readily found;
(iv)  The calculations are actually {\em easier}
      if the length of the system in one direction is infinite; and
(v)   The correlation length in this direction is
      simple to find since only the two largest eigenvalues of
      the TM are required.
For these reasons, there has recently been renewed
interest in calculating
interfacial properties by TM techniques.

Nevertheless, because the order of the TM increases exponentially
with the number of sites in a transverse layer
(discussed quantitatively in Sec.~\ref{sec-model}),
the size of systems which can be investigated through the
numerical diagonalization of a TM is greatly limited.
As a result, FSS analysis is even more important
for TM calculations than for Monte Carlo simulations.
The order of the TM has also limited
most previous attempts at calculating the
surface-tension anisotropy by TM techniques to
capillary-wave and SOS approximations \cite{MRS10,MRS11,Luck81}.
Specifically, in Refs.~\cite{MRS10,MRS11} Privman and
{\v S}vraki{\'c} derived FSS relations
for the correlation length of the Ising model in two and three dimensions
by using capillary-wave and SOS approximations;
these scaling relations
involve the surface-tension anisotropy.
The scaling relations for the 3D model were derived
through comparison with the 2D Ising model, which has been
solved exactly \cite{Onsager44,MRS12,MRS13,Ab91I} for
the boundary conditions used
in Refs.~\cite{MRS10,MRS11} to generate an
interface.  In the rough phase, a
longitudinal cross-section perpendicular to the interface was
compared with a 2D system, whereas in the smooth phase
the interface itself was compared with a 2D system.

In this article we present a large-scale numerical TM FSS study
of surface-tension anisotropy in 2D and 3D Ising models.
The analytic scaling relations of Privman and {\v S}vraki{\'c}
\cite{MRS10,MRS11} are applied to analyze free energies and
correlation lengths obtained by numerical diagonalization
of transfer matrices describing Ising models on both
\( \infty \! \times \! L \) square lattices and
\( \infty \! \times \! L \! \times \! M \)
cubic lattices \cite{MRS0} into
which an interface along the longitudinal ($x$)
direction can be introduced by applying appropriate boundary
conditions (discussed in Sec.~\ref{sec-model})
in the $y$-direction.
The purpose of this study is three-fold:
(i)   to test the analytically obtained scaling relations and
      possibly identify extensions or corrections;
(ii)  to investigate whether the limits on system size imposed by
      the present generation of parallel and vector supercomputers
      and diagonalization algorithms allow us to reach the asymptotic
      scaling regime in which FSS relations can yield information
      about interfacial properties in the thermodynamic limit;
(iii) to compare the resulting estimates of interfacial properties
      with estimates previously obtained
      by other analytical and numerical
      methods.
Throughout this study we use the
full Ising model, which includes contributions from
microstates with bubbles and overhangs as well as SOS microstates.

We now define the quantities to be studied in this work.
Given an interface (see Fig.~\ref{fig:3Dsys}) on average perpendicular
to the $y$-direction and making only a small angle $\theta$ with
the $xz$-plane, the
surface tension \(\sigma(\theta,T)\), measured per unit
projected area in the $xz$-plane, has a series
expansion for \( T \! > \! T_R \) \cite{MRS15}
\begin{equation}
   \label{eq:intro1}
   \sigma (\theta,T )/ \! \cos \theta =
   \tau(T) + {1\over 2} \kappa(T) \theta^2 + O(\theta^4) \; .
\end{equation}
The coefficients in this expansion are the surface free energy
\begin{equation}
   \label{eq:intro2}
   \tau(T) = \sigma (0,T) > 0
\end{equation}
and the surface stiffness coefficient
\begin{equation}
   \label{eq:intro3}
   \kappa(T) = \sigma (0,T) +
	\left.
		\frac{\partial^2 \sigma(\theta,T)}{\partial \theta^2}
	\right|_{\theta = 0}.
\end{equation}
If \(T \!<\! T_R,\)
$\kappa(T)$ is infinite, and the system can gain entropy only through
microscopic fluctuations, among which single-step terraces are
dominant.  Each single-step terrace contributes a quantity
$s(T),$ which is nonzero below $T_R$, to the
total free energy of the system; this quantity
is known as the step free energy.
The step free energy can also be related to the surface tension per
unit projected area by an expansion when
\(T \!<\! T_R\) \cite{PRF13}:
\begin{equation}
   \label{eq:intro4}
   \sigma (\theta,T )/ \! \cos \theta =
   \tau(T) + s(T) |\theta| + O(\theta^2) \; .
\end{equation}
These three
quantities, $\tau(T),$ $\kappa(T),$ and $s(T),$ are the
terms in which the anisotropy of the free energy will be discussed.

The remainder of this paper is organized as follows.
We define the model and discuss the appropriate
boundary conditions in Sec.~\ref{sec-model}.
In Sec.~\ref{sec-tau} we discuss estimates of the surface
free energy.
Section~\ref{sec-kappa} contains estimates of the surface
stiffness coefficient,
and Sec.~\ref{sec-step} is devoted to the step free energy.
In Sec.~\ref{sec-beta} we use the Roomany-Wyld approximant for the
Callan-Symanzik $\beta$ function \cite{Roomany,Barber}
to study the scaling of the correlation length, especially near the
roughening temperature.
Our results are summarized and discussed in
Sec.~\ref{sec-Concl}.

%%%%%%%%%%%%%%%%%%%%%%%    SECTION 2    %%%%%%%%%%%%%%%%%%%%%%
%%%%%%%%%%%%%%%%%%%%%%% Model & Methods %%%%%%%%%%%%%%%%%%%%%%%%

\section{Model and Methods}
  \label{sec-model}

We study the ferromagnetic Ising model with square and
cubic lattice geometries.
Each site $i$ has a spin \( s_i \! = \! \pm 1 \)
which interacts with the spin at each
nearest-neighbor site $j$ with an
interaction constant \( J_{i,j} \! > \! 0 \) .
The Hamiltonian for this system is given by
\begin{equation}
   \label{eq:model0}
	{\cal H} = - \sum_{\langle i,j \rangle}
		J_{i,j}  s_i s_j
		- \sum_{(i)} h_i s_i \;.
\end{equation}
Here \(\langle i,j \rangle\) denotes nearest-neighbor pairs
on a periodic lattice,
and $(i)$ represents boundary sites. For nearest-neighbor pairs
not on a boundary, \(J_{i,j} \! = \! J.\)  Throughout this
article the lattice constant is of unit length.

Various boundary conditions are needed to prepare both systems
with interfaces and systems without interfaces.
The presence or absence of an interface is determined by
conditions on the boundaries
\(y \! = \! 1\) and \(y \! = \! L.\)
All systems we study have the following:
(i)   a boundary interaction constant $J_y$ connecting the site at
      \( (x,y \! = \! 1,z) \) with the site at
      \( (x,y \! = \! L,z) \),
(ii)  a lower boundary magnetic field
      \(h_1\) applied on the \(y \! = \! 1\) plane, and
(iii) an upper boundary magnetic field
      \(h_L\) applied on the \(y \! = \! L\) plane.
Table~\ref{tab:bc}
summarizes the five types of boundary
conditions used in this paper.

It should be noted that the ``width'' of systems with
plus/plus and plus/minus
boundary conditions can be defined by measuring different
quantities.  These boundary conditions can be
imposed by coupling the planes at \(y \! = \! 1\)
and at \(y \! = \! L\) to planes of fixed spins.
In this case there are \(L \! + \! 1\) couplings in the
$y$-direction, and this number is often used
instead of the number $L$ of spin sites
as the measure of system width
(for example, in Refs.~\cite{MRS10,MRS11,Ab91I}).
In this article ``width'' refers to the
number of sites $L.$

A transfer matrix $\underline{T}$ can be used to find the free
energy per unit length in the longitudinal direction.
The largest eigenvalue of $\underline{T},$ \(\Lambda_0,\)
is related to the partition function $Z$ in the limit of
infinite longitudinal size ($K \! \rightarrow \! \infty$)
by \cite{MRS16,MRS17,MRS18,MRS19,MRS20}
\begin{equation}
   \label{eq:model1}
        \lim_{K \rightarrow \infty} Z^{1/K}
	= \lim_{K \rightarrow \infty}
		\left(
			\mbox{Tr} \left[ \underline{T}^K \right]
		\right)^{1/K}
	= \Lambda_0 \; .
\end{equation}
In our case the transfer matrix can be chosen to be
symmetric, taking into account both
the interactions within a transverse layer
and interactions between adjacent layers.
Using the thermodynamics of the canonical ensemble,
Eq.~(\ref{eq:model1}) can be used to find quantities such as
the free energy per site,
\begin{equation}
   \label{eq:model1a}
	f = \frac{-k_B T}{A} \ln \Lambda_0 \; ,
\end{equation}
where the area $A$ of a layer is given by
\( A \! = \! L \) in 2D and by \( A \! = \! LM \) in 3D.

For the Ising model, a layer in a
2D system consists of a 1D lattice
with $L$ sites in the $y$-direction, and in 3D a layer
is a 2D lattice with $L$ sites in the $y$-direction
and $M$ sites in the $z$-direction.
The transfer matrix $\underline{T}$
is a \( 2^A \times 2^A \) matrix.
Exact solutions have been found for the free energy and
correlation length in finite 2D systems
with antiperiodic boundary conditions \cite{MRS12}
or plus/minus boundary conditions \cite{MRS13,Ab91I},
so that systems with arbitrarily large values of $L$
can be used in FSS.
In 3D,  on the other hand, numerical diagonalization
of the large matrix $\underline{T}$ restricts us to
small values of $A$
due to limited computer memory. We are able to study
cross sections as large as \(A \! = \! 19\) on a
Cray Y-MP/432 by using the Numerical Algorithms Group ({\sc nag})
diagonalization routine F02FJF.
The largest cross section
we can study is \(A \! = \! 25\), for which an eigenvector of
$\underline{T}$ has $33\,554\,432$ elements.
Such large systems were run on a Thinking Machines Corporation
CM2 using a modified power algorithm \cite{Jen}
(for details see App.~\ref{app-cm2}).

The longitudinal correlation length is given by
by the ratio of the largest and next-largest eigenvalues
of $\underline{T}:$
\begin{equation}
   \label{eq:model3}
   \xi = 1 / \ln |\Lambda_0 / \Lambda_1| \; .
\end{equation}
Other length scales can be defined from smaller
eigenvalues of $\underline{T},$ and
scaling {\em Ans\"{a}tze} analogous to those we use
in the next two sections
have been proposed \cite{MRS10,MRS11}
for these length scales.  Here we consider
only the correlation length
defined by Eq.~(\ref{eq:model3}), since it is
expected that the other length scales require larger
values of $L$ (and in 3D, $M$)
to scale in the asymptotic fashion.

%%%%%%%%%%%%%%%%%%%%%      SECTION 3      %%%%%%%%%%%%%%%%%%%%%%
%%%%%%%%%%%%%%%%%%%%% Surface Free Energy %%%%%%%%%%%%%%%%%%%%%%

\section{The Surface Free Energy}
  \label{sec-tau}

In order to estimate the surface free energy from finite
systems, we compare the free energy of two
similar systems, one of which has an interface
and one of which does not. We attribute the
difference to the interface.
%  To estimate the surface free energy from finite
%  systems we compare the free energy of a system with
%  both an interface and bubbles with the free energy of
%  a similar system  containing only bubbles, and we attribute
%  the difference to the interface.
Specifically, we estimate
$\tau$ from antiperiodic ($a$) and periodic ($p$) boundary
conditions in the $y$-direction by \cite{MRS1a,MRS1b,MRS1c,MRS11}
\begin{equation}
   \label{eq:tau1}
   \tau^{a}_L (T)
	\equiv L \left(
		f_L^{(a)}(T) - f_L^{(p)}(T)
		 \right),
\end{equation}
and from plus/minus ($+/-$) and plus/plus (+/+) boundary conditions
in the $y$-direction by \cite{MRS1a,MRS1b,MRS1c,MRS11}
\begin{equation}
   \label{eq:tau2}
   \tau^{(+/-)}_L (T)
        \equiv L \left(
		f_L^{(+/-)}(T) - f_L^{(+/+)}(T)
		 \right).
\end{equation}
The factor $L$, which is the width of the system perpendicular to
the interface, is needed to convert the
difference between the volume densities $f$
into a surface density. These relations provide just one
possible definition of the surface tension in the 2D Ising model,
but it has been proven \cite{MRS22}
that all such definitions are equivalent with
that of Onsager for low temperatures.
These estimates are valid in any number of dimensions
(in 3D the subscript $L$ is replaced by $L,M$).

Figure~\ref{fig:t2D}a shows the 2D estimates
$\tau^{a}_L(T)$ defined in Eq.~(\ref{eq:tau1})
and the exact solution $\tau(T)$ in the infinite-lattice
limit \cite{Onsager44}.
The difference between the finite-size
estimate and the infinite limit
in this case is inversely proportional to
the correlation length in the periodic system, so that for
\(T \! < \! T_c,\) $\tau_L(T)$ converges to $\tau (T)$
exponentially with $L$ \cite{MRS12}.
Thus the size dependence is noticeable only near $T_c;$
in all cases, for 2D it has been shown  \cite{MRS23}
that the critical-point behavior is such that
\(\sigma(T)\) vanishes linearly with the reduced temperature,
\( t \equiv (T_c - T)/ T_c.\)

The 2D estimates $\tau^{(+/-)}_L(T)$ defined in Eq.~(\ref{eq:tau2})
are plotted in Fig.~\ref{fig:t2D}b.
In this case, however, we observe numerically that
the estimates from finite-width lattices
converge as a simple power law:
\begin{equation}
   \label{eq:tau3}
   \tau^{(+/-)}_L (T) - \tau(T) \approx A(T)L^{-B}.
\end{equation}
The effective scaling exponent is thus given by
\begin{equation}
   \label{eq:tau4}
   B(L,T) = - \frac{d \ln [\tau^{(+/-)}_L (T) - \tau(T)]}
	{d \ln L} \; ,
\end{equation}
where the derivative is calculated by multiple-point
finite differences.
By holding $T \! = \! T_c$ and calculating $\tau^{(+/-)}_L (T)$
for values of $L \! = \! 2^i \! \times \! 10,$
$i \! \in \! \{1,2,\ldots,15\},$ we can use
Aitken's \(\Delta^2\) method \cite{NumAnal}
to find $\lim_{L \rightarrow \infty}|B(L,T_c) - 1| < 10^{-9},$
in agreement with the prediction
$\lim_{L \rightarrow \infty}B(L,T_c) = 1$
made in Ref.~\cite{MRS20}.
Using the same method but fixing $k_B T = 1.5 J,$ we find that
for \(T \! < \! T_c,\)
$|\lim_{L \rightarrow \infty}B(L,T)-2| < 10^{-7}.$
For \(T \leq T_c\) and large $L$, the scaling exponent $B$ is
observed numerically to be a function only of the
critical scaling variable $tL^{1 / \nu} \! = \! tL$
(Fig.~\ref{fig:B}).

The estimates $\tau^{a}_{L,M}(T)$ for 3D are graphed in
Fig.~\ref{fig:t3D}a.
Again the convergence appears to be quite
rapid except near the Monte Carlo Renormalization
Group estimate \cite{MCRGTc} for $T_c.$
Similar rapid convergence is seen in the estimate
$\tau^{(+/-)}_{L,M}(T)$ for 3D (Fig.~\ref{fig:t3D}b).
The value of $\tau(T)$ is exactly known for 3D only at
\(T \! = \! T_c\) (where $\tau \! = \! 0$)
and at \(T \! = \! 0\) (where $\tau \! = \! 2J$),
but estimates from series expansions improved with
Pad{\'e} approximants \cite{FishPade} and from
Monte Carlo simulations \cite{Pinn92} agree well with our results
for $L \approx M.$
At $T_c,$ $\tau^a_{L,M}$
appears to converge to zero
as \(L^{-2}\) with little dependence on $M$, but it is difficult
to be certain with only a few small systems.  Recent analytic
results \cite{Morris92} for cubic ($K \! = L \! = \! M$)
systems with antiperiodic boundary conditions
indicate that the asymptotic scaling region is reached
only with very large systems.

%%%%%%%%%%%%%%%%           SECTION 4          %%%%%%%%%%%%%%%%%%
%%%%%%%%%%%%%%%% Surface Stiffness Coefficient %%%%%%%%%%%%%%%%%%

\section{The Surface Stiffness Coefficient}
  \label{sec-kappa}
%:::::::::::::::::::::%   SUBSECTION 1  %::::::::::::::::::::::%
%:::::::::::::::::::::%      d = 2      %::::::::::::::::::::::%
  \subsection{$\kappa$ in Two Dimensions}
    \label{ssec-k2d}
A capillary-wave calculation \cite{MRS10,MRS11}
yields that the correlation length for
antiperiodic boundary conditions and large $L$ should scale as
\begin{equation}
   \label{eq:k2d1}
   \xi^{a}_L(T) \approx \frac{2}{\pi^2} \frac{\kappa(T)}{k_BT} L^2
\end{equation}
for all \( T \! < \! T_c. \)
Solving this equation for $\kappa (T)$
gives a finite-size estimate for the
surface stiffness coefficient,
\begin{equation}
   \label{eq:k2d2}
   \kappa^{a}_L(T) = k_B T  \frac{\pi^2}{2} \frac{\xi^a_L(T)}{L^2}.
\end{equation}
Exact results \cite{Svrak88} agree with this relation
to leading order in $L$ \cite{MRS12}.
This estimate is shown
in Fig.~\ref{fig:k2D}a, as is the exact solution
in the infinite-lattice limit \cite{MRS12}.
We have analyzed the convergence and found that
for temperatures \( T \! \leq \! T_c \) the finite-lattice
estimate converges to its infinite-lattice
limit in a simple power-law fashion:
\begin{equation}
   \label{eq:k2d3}
   \kappa^{a}_L(T) - \kappa(T) \approx aL^{-b}.
\end{equation}
As in Sec.~\ref{sec-tau}, we use
multiple-point numerical differentiation
to determine an effective scaling exponent
\begin{equation}
	\label{eq:k2d4}
	b(L,T) = - \frac{d\ln [\kappa^a_L (T) - \kappa (T)]}
			{d\ln L}\;,
\end{equation}
which is displayed in Fig.~\ref{fig:b}a.
Although $b$ is not monotonic, it shares the most important
characteristics of the scaling exponent $B$ defined in
Eq.~(\ref{eq:tau4});
that is, $b$ varies from \( b \! = \! 1 \)
for \( T \! = \! T_c \) (in agreement with critical scaling)
to \( b \! = \! 2 \)
for \( T \! \ll \! T_c \), and $b$ is
a function only of the critical scaling variable $tL$
for large $L.$

The exact solution for
$[\xi_L^a (T)]^{-1}$ is a sum in which both the total number of
terms and each individual term depends on $L$ \cite{MRS12}.
We avoid the resulting difficulty of exactly
differentiating $\kappa_L^a$ with respect to $L$ by
using numerical differentiation in the evaluation of
Eq.~(\ref{eq:k2d4}).
However, it is also possible to define a ``reduced correlation length''
\(\hat{\xi}\), obtained
from the antiperiodic correlation length \(\xi^{a}\) and the
periodic correlation length \(\xi^{p}\) by
\begin{equation}
   \label{eq:k2d5}
   [\hat{\xi}_L(T)]^{-1} =
        [\xi^{a}_L(T)]^{-1} + [\xi^{p}_L(T)]^{-1},
\end{equation}
which contains only one explicitly $L$-dependent term
(see Eq.~(\ref{eq:bhat_6}) in App.~\ref{app-b}).
Replacing \(\xi^a_L (T)\) in Eq.~(\ref{eq:k2d2}) with
\(\hat{\xi}_L (T)\),
we find an estimate, \( \hat{\kappa}_L (T) \),
which is a differentiable function of $\ln (L)$ if $L$ is treated as
a continuous variable.
Except for small systems or $T \! \approx \! T_c,$
$[\xi^{p}_L(T)]^{-1}$ is vanishingly small in comparison
with $[\xi^{a}_L(T)]^{-1}.$
Thus it is clear that
\( \hat{\kappa}_L (T) \) will be less than
\( \kappa^a_L (T) \) but must converge to the same limit.
In fact, for all $L,$
\( \kappa(T) \! < \! \hat{\kappa}_L (T) \! < \! \kappa^a_L (T).\)
In the limit of large $L$ we find the effective scaling exponent
\begin{mathletters}
\begin{eqnarray}
   \hat{b}(tL) & = & 2 \left( \frac{1}
		    {1 - 4c(tL)^2[g(tL) - 2c]/\pi^2} \right)
	      \nonumber \\
   \label{eq:ap2d5}
	      & &    \times \left( 1 - \frac{\pi^2}
		    {2(tL)^2g(tL)[g(tL) - 2c]} \right) \; ,
\end{eqnarray}
where \(c = \sinh^{-1} (1) \approx 0.8814 \) and \( g(x) \)
is given by
\begin{equation}
   \label{eq:ap2d6}
   g(x) = \sqrt{ (2c)^2 + \left( \frac{\pi}{x} \right)^2 } \; .
\end{equation}
\end{mathletters}
Equation~(\ref{eq:ap2d5}) is derived in App.~\ref{app-b}.
As is the case for $B(tL)$
and $b(tL),$  $\hat{b}(tL)$ varies from
1 at $tL \! = \! 0$ to 2 at $tL \! = \! \infty,$ but unlike
$b(tL),$ $\hat{b}(tL)$ is monotonic (Fig.~\ref{fig:b}b).
It is proved in App.~\ref{app-b}
that $\lim_{tL \rightarrow \infty} \hat{b}(tL) = 2,$ and since
$\lim_{tL \rightarrow \infty} \hat{b}(tL) =
\lim_{tL \rightarrow \infty} b(tL),$ we also have
$\lim_{tL \rightarrow \infty} b(tL) = 2.$

Another capillary-wave calculation \cite{MRS10,MRS11}
indicates that for plus/minus boundary conditions
the correlation length scales as
\begin{equation}
   \label{eq:k2d7}
   \xi^{(+/-)}_L(T) \approx \frac{2}{3\pi^2}
	\frac{\kappa(T)}{k_BT} (L+2)^2
\end{equation}
for all \( T \! < \! T_c, \) giving the finite-size estimate for the
surface stiffness coefficient
\begin{equation}
   \label{eq:k2d8}
   \kappa_L^{(+/-)} (T) = k_B T \frac{3 \pi^2}{2}
			\frac{\xi^{(+/-)}_L(T)}{(L+2)^2} \;.
\end{equation}
Figure~\ref{fig:k2D}b shows \( \kappa^{(+/-)}_L(T) \).
We have observed numerically that for any fixed temperature
$T \! \leq \! T_c,$ $\kappa^{(+/-)}_L(T)$
converges to $\kappa(T)$ as $L^{-1}.$

Refs.~\cite{MRS10} and \cite{MRS11} actually state
Eqs.~(\ref{eq:k2d7}) for large $L$
with $\xi^{(+/-)}_L \! \propto \! L^2.$
However, in the same papers Privman and {\v S}vraki{\'c}
obtained estimates for the correlation length in 3D both for
$T \! > \! T_R$ (Eq.~(\ref{eq:k3d1b})) and for
$T \! < \! T_R$ (Eq.~(\ref{eq:step3})), which should be
related to the scaling relation of the correlation length in
2D (Subsec.~\ref{ssec-k3d}).  It is reasonable to suspect that
the $L$-dependence of the correlation length is not affected by
the roughening transition in the case of plus/minus boundary
conditions, just as it is not affected in the case of
antiperiodic boundary conditions (Eqs.~(\ref{eq:k3d1}) and
(\ref{eq:step1})).
This argument and our numerical observations lead us to conclude
$\xi_L^{(+/-)} \propto (L+2)^2.$

%:::::::::::::::::::::%   SUBSECTION 2  %::::::::::::::::::::::%
%:::::::::::::::::::::%      d = 3      %::::::::::::::::::::::%
  \subsection{$\kappa$ in Three Dimensions}
    \label{ssec-k3d}
There are two contributions to the correlation length along
the interface.  The first comes from pieces of interface which,
due to thermal fluctuations, are normal to the longitudinal
direction, which is the direction in which the
correlation length is measured. This
contribution is related to the surface stiffness $\kappa(T).$
The second contribution
comes from the formation of bubbles in the bulk and is much
less important for $T \! \ll \! T_c.$

Both of these contributions can be treated approximately by
considering a cross section of the 3D system
parallel to the $xy$-plane.  The relationship between
correlation length and surface stiffness in such a cross-section
should be the same as for a 2D system of the same width $L.$
Since there are $M$ such cross-sections, and since each contributes
on average the same amount of interface normal to the longitudinal
direction, the total effective surface stiffness for the 3D
system is given by $M\kappa(T)$ and has the same form as the
corresponding 2D systems; that is,
the appropriate estimate for the 3D surface
stiffness coefficient for antiperiodic boundary conditions
is given from Eq.~(\ref{eq:k2d2}) by
\begin{equation}
   \label{eq:k3d1}
   \kappa^a_{L,M}(T) = k_B T\frac{\pi^2}{2}
	\frac{\xi^a_{L,M}(T)}{L^2 M} \;,
\end{equation}
and for plus/minus boundary conditions from Eq.~(\ref{eq:k2d8}) by
\begin{equation}
   \label{eq:k3d1b}
   \kappa^{(+/-)}_{L,M}(T) = k_B T \frac{3\pi^2}{2}
	\frac{\xi^{(+/-)}_{L,M}(T)}{(L+2)^2 M} \;,
\end{equation}
for $L \! \approx M \! \gg \! 1.$
These estimates were proposed by Privman and
{\v S}vraki{\'c} \cite{MRS10,MRS11}.
No part of this argument
depends on the boundary conditions in the
$z$-direction, so Eqs.~(\ref{eq:k3d1}) and (\ref{eq:k3d1b}) are
valid for either periodic or free boundary conditions applied in the
$z$-direction.  In this article we discuss $\kappa(T)$ only
in the case of periodic boundary conditions in the
$z$-direction, but we have observed that little quantitative
change occurs when free boundary conditions are used.

In Fig.~\ref{fig:k3D}a the
``reduced correlation length,'' defined in 3D by analogy with
Eq.~(\ref{eq:k2d5}), is used in place of $\xi^{a}_{L,M}(T)$
to estimate $\kappa(T)$ using Eq.~(\ref{eq:k3d1}).
As in Subsec.~\ref{ssec-k2d}, this produces a
reduction of the estimate of $\kappa$ near $T_c$
(where $\kappa(T)$ is zero in an infinite system)
but cannot have any effect on the
estimate of $\kappa(T)$ for $T \! < \! T_c$ in the limit of large
systems.  Also shown in Fig.~\ref{fig:k3D}a are estimates of
$\kappa(T)$ based on series expansions improved by
Pad{\'e} approximants \cite{FishPade}
and estimates based on Monte Carlo simulations \cite{Pinn92}.
In spite of the small systems to which our study was limited,
we find good agreement between the
transfer-matrix estimates in this study and the previous
estimates \cite{FishPade,Pinn92}
over a wide range of temperatures between $T_R$ and $T_c.$

In Fig.~\ref{fig:k3D}b we show the estimate for $\kappa(T)$
that results from using Eq.~(\ref{eq:k3d1b}).  Comparison
with the series expansion and Monte Carlo estimates
shows that finite-size corrections to Eq.~(\ref{eq:k3d1b})
cannot be neglected at any temperature
for the systems we studied.

The surface stiffness coefficient for the infinite system
shows a characteristic KT discontinuity \cite{KT73,KT74} at $T_R,$
jumping from \(\kappa(T) = \infty \) for all \(T \! < \! T_R\)
to \(\kappa(T)/k_BT = \pi/2 \) for \(T \! = \! T^{+}_R\)
\cite{Wolf85}.
If we use the criterion \(\kappa(T_R)/k_B T_R = \pi/2 \) to estimate
$T_R$ while using antiperiodic boundary conditions,
we find that the estimated value of $T_R$ depends most strongly
on the value of $M$ (Fig.~\ref{fig:Tr}).
For the system $(L,M) = (3,8),$ we find
$T_R \approx 2.337 J/k_B.$
This represents a deviation of approximately
5\% from the Monte Carlo estimates \cite{HasenPhD,MRS5,MCRGTc},
and is comparable with our alternative estimates of $T_R$
discussed in Sec.~\ref{sec-beta}.
Although the convergence in Fig.~\ref{fig:Tr} appears to
be roughly power-law in behavior, much slower convergence
should be expected for large systems due to the
KT nature of the roughening transition \cite{Nijs82}.

%%%%%%%%%%%%%%%%%%%%%%     SECTION 5    %%%%%%%%%%%%%%%%%%%%%%
%%%%%%%%%%%%%%%%%%%%%% Step Free Energy %%%%%%%%%%%%%%%%%%%%%%

\section{The Step Free Energy}
  \label{sec-step}

For \( T \! < \! T_R, \) a SOS calculation
indicates that the correlation length
for antiperiodic boundary conditions should scale
as \cite{MRS10,MRS11}
\begin{equation}
	\label{eq:step1}
	\xi^a_{L,M}(T) \approx  \mu(T) M^w
	\exp \left[ \frac{M s(T)}{k_B T} \right]
	\left(\sin \frac{\pi}{2L} \right)^{-2},
\end{equation}
where \(\mu(T) \equiv \frac{1}{4}\exp[wC(T,L/M)],\)
and \(C(T,L/M)\) is a function possibly of the temperature,
boundary conditions, and shape of the layer, but is independent
of the size of the layer.
The boundary conditions in the $z$-direction may be either
periodic (Fig.~\ref{fig:s}a) or free (Fig.~\ref{fig:s}b).
By taking three different values of $M$ we can solve
simultaneously for \(s(T),\)
\(w(T),\) and \(\mu(T).\)
The step free energy $s(T)$ discussed in this article is the
step free energy for an interface parallel to the
$xz$-plane. A low-temperature expansion for the step free energy
of inclined planes can be found in Ref.~\cite{Holzer89}.

Unfortunately, numerical convergence difficulties
limit to very small systems our studies of interfaces
at very low temperatures.
Furthermore, quantities near $T_R,$
where we can use larger systems,
may be expected to be difficult to estimate due to the KT transition.
Nevertheless, the estimates of $s(T)$ rapidly converge
at low temperatures,
so that the limitation to small systems may not present
difficulties in estimating the
step free energy. Also, the estimates
of $s(T)$ are rather insensitive to the values of $w$ and
$\mu(T),$ so that problems at low temperatures
with the last two quantities (described below)
do not seem to create difficulties in the estimates of $s(T).$
This is because the divergence of the correlation length at
low temperatures is dominated by the exponential factor in
Eq.~(\ref{eq:step1}), which contains the step free energy.

The estimates of $s(T)$ derived from Eq.~(\ref{eq:step1})
are compared in Fig.~\ref{fig:s} with approximations
valid either near $T \! = \! 0$ or near $T \! = \! T_R.$
At low temperatures the interface becomes a series
of plateaus separated in height from their neighbors
by one lattice spacing.  These plateaus are equivalent
to domains in a 2D system \cite{Holzer89,MRS11};
the equivalence is exact at $T \! = \! 0.$
This yields the well-known approximation \cite{Holzer89}
\begin{equation}
	\label{eq:step2}
	s(T) \approx \tau^{\rm 2D}(T)\;.
\end{equation}
We use $\tau^{\rm 2D}(T)$ as an approximation
to $s(T)$ for $T \! \ll \! T_R$ for the following reasons:
it provides a simple physical picture;
it agrees with the first several terms of
   low-temperature series expansions of $s(T)$ \cite{Holzer89};
it has been solved exactly \cite{Onsager44}; and
the 2D critical temperature is a rigorous lower bound on
   $T_R$ \cite{Beij75}.
At higher temperatures $(T \! \approx \! T_R),$
$s(T)$ shows the characteristic KT essential singularity
at $T_R$ \cite{Beij77,KT73,KT74}, and is given by
\begin{equation}
	\label{eq:beta2}
	\frac{s(T)}{k_BT} \approx \exp
		 \left(
			-\frac{\pi}{2c}
			\sqrt{\frac{T_c}{T_R - T}} \;
		\right) \;,
\end{equation}
where the nonuniversal parameter $c$ has been found by
Monte Carlo simulations \cite{MRS4,F92PRL} to be
$c \! = \! 1.57 \pm 0.07.$ In Fig.~\ref{fig:s} we have calculated
the estimate Eq.~(\ref{eq:beta2}) with $c \! = \! \pi/2.$

The correspondence between a low-temperature interface in 3D and
a 2D system ensures \cite{Priv83,Cabrera86,Cabrera87,Barber87}
that for periodic boundary conditions in
the $z$-direction \(w \! = \! 1/2,\) whereas for
free boundary conditions in
the $z$-direction \(w \! = \! 0.\)
Our estimates, however, have \(w \! = \! 0\) for
sufficiently low temperatures for both sets of boundary conditions
(Fig.~\ref{fig:w}).
The incorrect estimate for
$w$ in the case of periodic boundary conditions in the
$z$-direction provides a caution that the asymptotic scaling
regime, in which corrections to Eq.~(\ref{eq:step1})
can be expected to be very small,
has not been reached with the current small systems.
The parameter $\mu(T)$ (not shown) is consistent with
$w \! = \! 0$ at low temperatures.

For plus/minus boundary conditions, the scaling relation
\mediumtext
\begin{equation}
	\label{eq:step3}
	\xi^{(+/-)}_{L,M}(T) \approx  \mu(T) M^w
	\exp \left[ \frac{M s(T)}{k_B T} \right]
	\left[
	      \sin \frac{3\pi}{2(L+2)}
	      \sin \frac{\pi}{2(L+2)}
	\right]^{-1}
\end{equation}
\narrowtext
is expected in analogy with Eq.~(\ref{eq:step1}) on the basis
of another SOS calculation \cite{MRS10,MRS11}.
The analysis for Eq.~(\ref{eq:step3}) parallels
that for Eq.~(\ref{eq:step1}),
so that at zero temperature we should find that
for periodic boundary conditions in the $z$-direction
\(w \! = \! 1/2,\) and that for free boundary conditions
\(w \! = \! 0.\)
In this case the values extracted
for $s(T)$ (not shown) for different system sizes are inconsistent
with each other even at low temperatures,
and the corresponding values of $w$ (also not shown)
show no pattern whatsoever.

In short, the scaling relation of
Eq.~(\ref{eq:step1}) yields values of the step free energy
that are consistent with the low-temperature
approximation $\tau^{\rm 2D}(T),$ whereas
Eq.~(\ref{eq:step2}) does not even yield consistent estimates
at low temperatures.
The parameter $w(T)$ is found to be wrong at low temperatures
when calculated from Eq.~(\ref{eq:step1}) with periodic boundary
conditions in the $z$-direction, and when calculated from
Eq.~(\ref{eq:step2}) it yields inconsistent values.
It is thus apparent that large corrections
to scaling are present for the systems in this study.

%%%%%%%%%%%%%%%%%%%%%%     SECTION 6    %%%%%%%%%%%%%%%%%%%%%%
%%%%%%%%%%%%%%%%%%%%%%  Beta Function   %%%%%%%%%%%%%%%%%%%%%%

\section{The Callan-Symanzik $\beta$ Function in the Rough Phase}
  \label{sec-beta}

In Subsec.~\ref{ssec-k3d} and Sec.~\ref{sec-step} we have
evaluated terms in the surface tension from
FSS relations for the correlation length
\cite{MRS10,MRS11} in 3D for $0 \! < T \! < \! T_c.$
As discussed in Subsec.~\ref{ssec-k3d},
for $T_R \! \leq \! T \! \leq \! T_c$ each of the $M$
cross-sections parallel with the $xy$-plane makes a contribution
to the correlation length in an
$\infty \! \times \! L \! \times \! M$
system which is of the same form as the correlation length
in a 2D system of width $L.$
{}From this argument we have Eqs.~(\ref{eq:k3d1}) and (\ref{eq:k3d1b}),
which give $\xi_{L,M} \! \propto \! M$ for either antiperiodic or
plus/minus boundary conditions.
Such linear divergence with system size of the correlation length
is characteristic of a critical point \cite{MRS20},
and indicates that the
rough phase of the interface is a critical phase \cite{RoughRev}.
A quantity that has proven particularly useful in locating
such extended critical phases in previous studies
({\em e.g.} Refs.~\cite{Kitatani,Dunweg}) of systems
with Kosterlitz-Thouless behavior is the Roomany-Wyld
approximant to the Callan-Symanzik $\beta$ function
\cite{Roomany,Barber}.
This approximant is defined by
\begin{equation}
	\label{eq:beta1}
	\beta_{L,M}(T) = \frac{(d \ln [\xi_{L,M}/L^2])/(d \ln M) - 1}
	                      {(d \ln [\xi_{L,M}/L^2])/(d T)} \;,
\end{equation}
where the derivative with respect to $\ln M$ is evaluated by
multiple-point finite differences.
(There should be no confusion between $\beta_{L,M}(T)$ and the
dimensionless inverse temperature,
which is called $\beta$ in the appendices.)
Within the region where a ``critical interface'' exists,
one should have $\beta_{L,M}(T) \! = {\rm const} \! \approx \! 0.$
In this section we apply the $\beta$ function to the
3D Ising model with antiperiodic boundary conditions in order
to obtain independent TM FSS estimates both for $T_R$ and
for the (size-dependent) temperature where the
rough-interface phase is
destroyed by critical-point fluctuations in the small systems we
are studying.

In Fig.~\ref{fig:beta} we show various estimates of $\beta_{L,M}(T)$
for the case of antiperiodic boundary conditions in the $y$-direction
and periodic boundary conditions in the $z$-direction. Finite-size
effects were larger for the case of free boundary conditions
in the $z$-direction and much larger for the case of
plus/minus boundary conditions in the $y$-direction, so
these results are not shown.
The dashed curve near $T \! = \! 0$ is the value of
$\lim_{L,M \rightarrow \infty} \beta_{L,M}(T)$ where
$\tau^{\rm 2D}(T)$ is substituted for $s(T)$ in
Eq.~(\ref{eq:step1}). If the scaling relations
of Eqs.~(\ref{eq:step1}) and (\ref{eq:step3}) for $T \! < \! T_R$
are to join continuously with the scaling relations
of Eqs.~(\ref{eq:k3d1}) and (\ref{eq:k3d1b}) for $T \! > \! T_R$
at $T_R,$ then $w(T_R) \! = \! 1.$
The solid curve for $T \! \leq \! T_R$ traces
$\lim_{L,M \rightarrow \infty} \beta_{L,M}(T)$
under the assumptions that
Eq.~(\ref{eq:step1}) holds near $T_R$ with $w \! = \! 1,$ and that
the step free energy near $T_R$ is given by Eq.~(\ref{eq:beta2}).
The agreement of the data in Fig.~\ref{fig:beta} with this
portion of the solid curve is good despite the fact that
Eq.~(\ref{eq:step1}) is based on the assumption that the
terraces in the interface are well separated, which is not
the case near $T_R.$
The solid horizontal line $\beta_{L,M}(T) \! = \! 0$ for the
temperature range $T_R \! \leq \! T \! \leq \! T_c$ shows
$\lim_{L,M \rightarrow \infty} \beta_{L,M}(T)$
assuming that Eq.~(\ref{eq:k3d1}) holds
in this range.

The finite-size estimates for $\beta_{L,M}(T)$ are of two types.
The first type consists of evaluating
$(d \ln [\xi_{L,M}/L^2])/(d \ln M)$ by a three-point estimate
while holding $L$ constant,
and then making the empirical extrapolation
\begin{equation}
	\label{eq:beta3}
	\beta_{L,(M \rightarrow \infty)}(T)
		= \beta_{L,M}(T) + a(T)\exp[-b(T)M]
\end{equation}
where $a(T)$ and $b(T)$ are unknowns that depend only
on temperature.
(The only justification for this extrapolation procedure
is that it appears to work nicely
for small systems, although we expect
logarithmic convergence for large systems because
roughening is a KT transition \cite{Nijs82}.)
For $L \! > \! 3$ this yields results which depend little on
$M$ and show the emergence of a plateau near $T_R$ with
$\beta_{L,M}(T) \! \approx \! -0.06.$
Even for $L \! = \! 5,$ critical finite-$L$ effects destroy the
plateau for temperatures greater than about $3.2 J/k_B.$
The extrapolated value of
$\beta_{L,(M \rightarrow \infty)}(T)$ for $L \! = \! 3$ does
not show any plateau, but there is a change in concavity
at about $2.41 J/k_B,$ where the
critical finite-$L$ effect gives way to
the $T \! < \! T_R$ behavior.

The second type of finite-size estimate is made by
using only systems with square cross-sections
$(L \! = \! M)$ in the numerical evaluation
of $(d \ln [\xi_{L,M}/L^2])/(d \ln M)$ in Eq.~(\ref{eq:beta1}).
This estimate of $\beta_{L,M}(T)$
yields a plateau at $\beta_{L,M}(T) \! \approx \! -0.1.$

In either case, the onset of the plateau (which corresponds to
an estimate of $T_R$) can be defined by a peak in
$d^2 \beta_{L,M}(T) / dT^2.$  This estimate yields
$T_R \approx 2.3 J/k_B,$ which
is approximately the same as
that given in Subsec.~\ref{ssec-k3d}.
The high-temperature end of the plateau seems to correspond to
the temperature at which critical finite-$L$ effects become large
in the interface internal energy (not shown), which
is given by
\begin{equation}
	\label{eq:beta4}
	U = -T^2   \frac{\partial (\tau/T)}
			{\partial     T   } \;.
\end{equation}

%%%%%%%%%%%%%%%%%%%%        SECTION 7        %%%%%%%%%%%%%%%%%%
%%%%%%%%%%%%%%%%%%%%  Comments & Conclusion  %%%%%%%%%%%%%%%%%%

\section{Discussion and Conclusion}
  \label{sec-Concl}

As a nonperturbative numerical method for studying semi-infinite
systems in equilibrium, numerical transfer-matrix (TM)
calculations followed by finite-size scaling (FSS) \cite{MRS10,MRS11}
provide an appealing alternative
to the analytical and Monte Carlo methods
which have been used to study interfaces in the Ising model.
Nevertheless, the rapid growth of the order of the transfer matrix
with system size has limited most previous studies to
Solid-on-Solid (SOS) and capillary-wave approximations.

In this paper we have performed
large-scale numerical TM calculations on a Cray Y-MP/432
and a Thinking Machines Corporation CM2
to obtain the correlation lengths and free energies of 2D and
3D Ising ferromagnets with interfaces, which are related to
the surface tension anisotropy through the scaling relations
derived in Refs~\cite{MRS10,MRS11}.
In 2D we are able to analyze the corrections to scaling.
In 3D we are only able to confirm the scaling relations for
antiperiodic boundary conditions, and the resulting estimates
for $\tau(T)$ and $\kappa(T)$ agree well with estimates
from series expansions improved with Pad{\'e}
approximants \cite{FishPade} and from
Monte Carlo simulations \cite{Pinn92}.
For plus/minus boundary conditions the finite-size effects
in our estimates
are still large, even for the largest systems which can be studied
using current supercomputers and algorithms, and the asymptotic
scaling regime has not been reached.

We find that although in 2D
the surface free energy converges exponentially
for antiperiodic boundary conditions, for plus/minus boundary
conditions the convergence obeys a power law.  The effective scaling
exponent for plus/minus boundary conditions is a function only of
the critical scaling variable $tL,$
and varies monotonically from 1 at $tL \! = \! 0$ to
2 as $tL \! \rightarrow \! \infty.$
Although the exact value of $\tau(T)$ is not known in 3D
for arbitrary temperatures,
our numerical results for $L \! = \! M$
agree well with the estimates of Refs.~\cite{Pinn92,FishPade}.
We are not able to
make a quantitative study of the scaling.

In 2D the surface stiffness coefficient is found from
scaling relations based on capillary-wave studies
\cite{MRS10,MRS11}.
For antiperiodic boundary conditions we find that
$\kappa_L^a(T)$ at any fixed temperature $T \! < \! T_c$
converges in a power-law fashion with an effective
exponent similar to that found for $\tau^{(+/-)}(T).$
Specifically, the exponent is a function of
the critical scaling variable $tL,$
and varies from 1 at $tL \! = \! 0$ to
2 as $tL \! \rightarrow \! \infty,$ though not monotonically.
By using the ``reduced'' correlation length, a monotonic
exponent, discussed in App.~\ref{app-b}, can be calculated in
the infinite-lattice limit.
For plus/minus boundary conditions
$\kappa_L^{(+/-)}(T)$ converges simply as $1/L.$

The estimates for $\kappa(T)$ in 3D are based on scaling relations
similar to those in 2D because of the similarity of
a cross-section in 3D to a 2D system.
These estimates \cite{MRS10,MRS11}
in the case of antiperiodic boundary conditions
agree well with both series \cite{FishPade}
and Monte Carlo  \cite{Pinn92} estimates.
Plus/minus boundary conditions lead to large
finite-size effects in the estimates of $\kappa(T)$ at all
temperatures.
Estimates for the roughening temperature based on
the KT relation $\kappa(T_R)/k_B T_R^+ \! = \! \pi/2$
\cite{KT73,KT74,Wolf85}
depend more strongly on the value of $M$ than on the value of $L.$
For $(L,M) \! = \! (3,8),$ this estimate
yields $T_R \! \approx \! 2.3.$

Solid-on-Solid arguments have yielded
scaling relations for the correlation length
\cite{MRS10,MRS11} which we exploit to find $s(T)$.
In the case of antiperiodic boundary conditions,
the estimates for $s(T)$ agree well with other approximations at
low temperatures.  The low-temperature value of the fitting
parameter $w$ for systems with periodic boundary conditions
in the $z$-direction, however, is in contradiction
with known results \cite{Priv83,Cabrera86,Cabrera87,Barber87}.
This indicates the presence of corrections to scaling for
the correlation length in the small systems of this study.
These corrections are even more pronounced in the case of
plus/minus boundary conditions, where the values of
$s(T)$ estimated from different system sizes are inconsistent
even at low temperatures.

The rough phase of the interface,
which is a type of critical phase \cite{RoughRev},
displays scaling behavior for the correlation length
which is similar to that of other critical systems \cite{MRS20}.
Specifically, as a result of Eq.~(\ref{eq:k3d1}),
we have $\xi \! \propto \! M.$
We used the Roomany-Wyld approximant to search for a plateau in
the Callan-Symanzik $\beta$ function \cite{Roomany,Barber} which
corresponds to this critical temperature range.  The onset of the
plateau agrees roughly
with the estimate of $T_R$ from Subsec.~\ref{ssec-k3d},
while the end of the plateau corresponds to the onset of
large-scale finite-size corrections in the interface internal energy
caused by the disappearance of the interface at $T_c.$

In summary, we have provided transfer-matrix finite-size
scaling estimates of the quantities which characterize the
surface-tension anisotropy in two- and three-dimensional
Ising ferromagnets: the surface free energy $\tau(T),$
the surface stiffness coefficient $\kappa(T),$ and the
step free energy $s(T).$  Our results support the
analytically obtained relations of Refs.~\cite{MRS10,MRS11}
and are fully consistent with existing series and Monte Carlo
estimates \cite{Pinn92,FishPade}.  However, we find that in the
three-dimensional case the limitations of current computer
technology and algorithms do not permit transfer-matrix
studies of sufficiently large systems to improve on
the numerical accuracy of the previously existing estimates.

%..................... ACKNOWLEDGEMENTS .....................%

	\acknowledgements

The authors wish to thank K.~K.~Mon and V.~Privman
for useful discussions, M.~Grant for a reading of
the manuscript, and P.~E.~Oppenheimer
for discussions about PARIS programming of the CM-2.
The authors would also like to thank
M.~E.~Fisher for assistance in locating relevant research and
for providing unpublished independent series estimates from his work with
H.~Wen, and would like to thank M.~Hasenbusch and K.~Pinn for
providing unpublished independent Monte Carlo estimates for
comparison.
This research was supported in part by the Florida State University
Supercomputer Computations Research Institute which is partially funded
by the U.S. Department of Energy through contract \# DE-FC05-85ER25000,
by the FSU Center for Materials Research and Technology, and
by the National Science Foundation through grant \# DMR-9013107.

%%%%%%%%%%%%%%%%%%%%        APPENDIX 1        %%%%%%%%%%%%%%%%%%
%%%%%%%%%%%%%%%%%     Computational Details     %%%%%%%%%%%%%%%%

\appendix{Computational Details}
   \label{app-cm2}
This appendix contains details about the algorithms used in the
diagonalization and the programming implementation of these algorithms.
In this appendix we take $k_B \! = \! J \! = 1,$ so that all
quantities are dimensionless.

If each of the identical layers has $A$ spin sites,
the $2^A\times 2^A$ transfer matrix
$\underline T$ in Eq.~(\ref{eq:model1}) can be decomposed into the
product of two matrices, $\underline T = \underline A \underline D$.
The matrix $\underline D$, which can be chosen to be diagonal, takes into
account both interactions between spins in the same layer and interactions
of the spins with a magnetic field.  The matrix $\underline A$ takes into
account the interaction between one layer and the next layer.
It is a direct
(Kronecker) product of $A$ identical $2\times2$ matrices
\begin{equation}
\label{eq:2x2}
{\underline a} = \left ( \matrix{
{\rm e}^\beta     & {\rm e}^{-\beta} \cr
{\rm e}^{-\beta}  & {\rm e}^\beta    \cr
}\right ) ,
\end{equation}
where $\beta \! = \! 1/T.$
In both of the algorithms used, as described below, the core of the
algorithm demands that for any given vector $\vec v$ one can obtain
a vector $\vec w = \underline T \vec v$.  In order to be able to
find even a few of the largest eigenvalues of $\underline T$
with $A$ large, one needs to minimize the
computer storage required.  This means that only a few
(typically between 2 and 20) vectors of size $2^A$ can be
stored, whereas the entire $2^{2A}$ elements of $\underline T$ cannot be
stored.  The matrix $\underline D$ only requires storage of $2^A$
elements, so it is computed once for each temperature and stored.
All of the boundary conditions
used in this paper are incorporated into the matrix $\underline D$.
Two slightly different methods of computing $\underline A \vec x$
were used.

Although the size of the matrix $\underline T$ can be reduced by
consideration of the symmetries of the model,
the savings in computer storage are not great in systems into
which an interface, which breaks many symmetries, has been introduced.
Furthermore, the symmetry-reduced matrix would reduce the speed
of the diagonalization program discussed below on the CM-2
by complicating interprocessor communications.
Since very few systems with
reasonable aspect ratios could be added by using symmetry
reduction, and because it would be inefficient to implement
symmetry reduction on the systems now accessible, we have not
pursued this.

For $A \! \le \! 19,$ the NAG routine F02FJF
was used on the FSU YMP, which has 32 MWords of memory.
This routine required use of the symmetric transfer matrix
$\underline D^{1/2} \underline A \underline D^{1/2}$.
One can write
${\underline A} = {\underline A}_A {\underline A}_{A-1} \cdots
{\underline A}_1$, where each $2^A \! \times \! 2^A$ matrix
${\underline A}_i$ is a direct product of
$A \!- \!1$ identity matrices of size $2 \! \times \! 2$
and the matrix ${\underline a}$ in position $i$.
A permutation matrix ${\underline P}$ can be defined such that
${\underline P}^{-1} {\underline A}_{i} \> {\underline P}
= {\underline A}_{i+1}$ with
${\underline A}_{A+1} = {\underline A}_1.$
In other words, $\underline P$ permutes the matrices
within the Kronecker product in a cyclic fashion
\cite{re:NovSuz,re:MAN1,re:MAN2,re:MAN3a,re:MAN3b,re:Audit91}.
For any $j,$ it follows that
${\underline A}_j = {\underline P}^{-j+1} {\underline A}_1
{\underline P}^{j-1},$
and thus one obtains
$\underline A = {\underline P}^{-A+1} {\underline A}_1
({\underline P}{\underline A}_1)^{A-1}.$
Since ${\underline P}^{-A}$ is the identity matrix,
all one needs to do is
to program the multiplication of a vector by
${\underline A}_1$ and by ${\underline P}$ and repeat these $A$ times to
obtain the multiplication by $\underline A$.  The matrix
${\underline A}_1$ has only two non-zero elements in each row and column,
so multiplication by ${\underline A}_1$ can be efficiently programmed.
One way of doing this is to break the
vector to be multiplied into an upper and a lower part, each having
$2^{A-1}$ elements.  Then
\begin{equation}
\label{eq:Mult1}
{\underline A}_1 \pmatrix{{\vec v}_a \cr {\vec v}_b } =
{\rm e}^\beta \pmatrix{{\vec v}_a \cr {\vec v}_b }  +
{\rm e}^{-\beta} \pmatrix{{\vec v}_b \cr {\vec v}_a }.
\end{equation}
%
%                                          + + + + - - - -
%                                          + + - - + + - -
%                                          + - + - + - + -
%                                    +++ ( 1 0 0 0 0 0 0 0 )
%                                    ++- ( 0 0 0 0 1 0 0 0 )
%                        { J I   }   +-+ ( 0 1 0 0 0 0 0 0 )
%               Here P = {   J I } = +-- ( 0 0 0 0 0 1 0 0 )
%                        { I   J }   -++ ( 0 0 1 0 0 0 0 0 )
%                                    -+- ( 0 0 0 0 0 0 1 0 )
%                                    --+ ( 0 0 0 1 0 0 0 0 )
%                                    --- ( 0 0 0 0 0 0 0 1 )
%
The matrix $\underline P$ acts on a vector as
${\underline P} { {\vec v}_a \choose {\vec v}_b }
= \vec w,$ with
the $i$th even element of $\vec w$ given by the
$i$th element of ${\vec v}_b$ and
the $i$th odd element of $\vec w$ given by the
$i$th element of ${\vec v}_a$.
The multiplication by ${\underline P}$ can be efficiently
programmed using scatter-gather routines on a vector computer.

For $14 \! \le \! A \! \le \! 25,$ a double-precision program
in FORTRAN/PARIS for a
Thinking Machines Corporation massively parallel CM-2 computer was
used for the diagonalization.  The use of the PARIS programming language
was essential for diagonalization of the large matrices, since
the FORTRAN compiler would have otherwise
allocated more than the minimum number of arrays.
Programming simplicity in the PARIS portion of the program
required that the number of 1-bit
processors used be less than or equal to the size of a single vector.
The CM-2 available has $2^{16}$ processors, and can be divided into
quarters with $2^{14}$ processors each, giving the limit
$14 \! \le \! A.$
The upper limit was imposed by the computer memory on the CM-2.
To calculate the
correlation length, at least two vectors need to be iterated.
This entailed storing 6 double-precision (64-bit) vectors:
one of these a work vector, one the vector associated with the
diagonal matrix $\underline D$, and two vectors each for the
current iteration vector and for the next iteration vector.
In addition, $A \! - \! 1$ single-bit vectors were used to efficiently
implement the multiplication of a vector by the matrices
${\underline A}_i$.  (Temporary vectors used in the formation of
the bit vectors and of the diagonal matrix were discarded before the
64-bit iteration vectors were allocated.)
For $A \! = \! 25,$ this required use of $1.7$~Gbytes
of the available $2$~Gbytes of main memory on the CM-2.
The implementation
of the multiplication by one of the $A$ matrices
${\underline A}_i$ was made using the
hypercube communication of the CM-2.
This was done by using the decomposition
\( {\underline A}_i = {\rm e}^\beta {\underline I}
+ {\rm e}^{-\beta} {\underline X}_i.\)
Here \(\underline I\) is the \(2^A \! \times \! 2^A\)
identity matrix, and the permutation matrix
${\underline X}_i$ is a direct product of $A \! - \! 1$
identity matrices of size $2 \! \times \! 2$ and the matrix
${\underline X} = {0 1 \choose 1 0}$ in the $i$th
position.
The multiplication by the ${\underline X}_i$ matrices was performed
using efficient communication with the PARIS command
GET\_FROM\_POWER\_TWO, which allows hypercube communication to elements
a power of two away.

The program was written to diagonalize any real matrix
with both the largest eigenvalues
and corresponding eigenvectors real.
The algorithm used is based on a generalized power method
and proceeds as follows \cite{re:WILKINSON}.
A $2^A \! \times \! R$ matrix $\underline U$
is initialized with random elements.
Due to constraints of the PARIS program, $R$ must be
an integer power of 2, and memory constraints dictate
that $R \! = \! 2$ if $A \! = \! 25.$
Then the $2^A \! \times \! R$ matrix
$\underline V = \underline T \underline U$ is obtained using
the multiplication procedure described above for each of the
$R$ vectors in $\underline V.$
Two $R \! \times \! R$ matrices
{$\underline {\cal S} = {\underline U}^T {\underline U}$
and ${\underline {\cal Q}} = {\underline U}^T {\underline V}$
are then formed.
The eigenvalue equation
${\underline {\cal S}^{-1}} {\underline {\cal QG} }
= {\underline {\cal GD}}$ is then solved to find the
$R \! \times \! R$ diagonal matrix of eigenvalues
$\underline {\cal D}$ and the $R \! \times \! R$ orthogonal
matrix of normalized eigenvectors $\underline {\cal G}.$
The diagonal elements of $\underline {\cal D}$ provide
the estimates for the eigenvalues of the transfer matrix
${\underline T}$ for this iteration.
The matrix $\underline U$ is then updated
by ${\underline U} \leftarrow {\underline {{\rm V}{\cal G}} },$
and the process is repeated.

Near $T_c$ typically about 25 iterations were necessary to obtain
convergence of the correlation length to one part in $10^{15}$.
However, at lower temperatures many more iterations were required for
convergence.  Near $T_c$ for $A=25,$ the computer
time required for convergence on the $2^{16}$ processor CM-2
with double-precision Weitek floating-point accelerators was
about 15~minutes.

%%%%%%%%%%%%%%%%%%%%        APPENDIX 2        %%%%%%%%%%%%%%%%%%
%%%%%%%%%%%%%%%%%%%%     Calculation of b     %%%%%%%%%%%%%%%%%%

\appendix{The Scaling Exponent for $\hat{\kappa}$ in 2D}
   \label{app-b}

In Eq.~(\ref{eq:ap2d5}) in Subsec.~(\ref{ssec-k2d})
we presented the exact result for
an effective scaling exponent $\hat{b}(L,T)$
such that
\begin{equation}
	\label{eq:bhat_0}
	\hat{\kappa}_L (T) - \kappa(T) \approx
		\hat{a} L^{-\hat{b}} \; .
\end{equation}
Then, in analogy with Eq.~(\ref{eq:k2d4}),
\begin{eqnarray}
	\label{eq:bhat_1a}
	\hat{b}(L,T) & \equiv &
	   \frac{d\ln [\hat{\kappa}_L (T) - \kappa (T)]}{d\ln L} \\
	& = &
	\label{eq:bhat_1}
	-\frac{L}{\hat{\kappa}_L (T) - \kappa (T)}
	               \frac{d\hat{\kappa}_L (T)} {dL}\;.
\end{eqnarray}
This appendix contains a sketch of a derivation
of Eq.~(\ref{eq:ap2d5}), which is made by
rewriting the functions in
Eq.~(\ref{eq:bhat_1}) in terms of the reduced temperature
\( t \equiv (T_c - T)/ T_c\)  and the critical scaling
variable \(x = tL\), and then keeping only the
$O(1)$ term in $t$ at fixed $x.$ In this appendix we take
$k_B \! = \! J \! = \! 1$ and
measure the surface stiffness coefficient
in units of $k_B T,$ so that all quantities are dimensionless.

First, we define the functions which are to be expanded.
These are \cite{MRS10,MRS11,MRS12}
the inverse temperature
\begin{equation}
	\label{eq:bhat_2}
	\beta = 1/T\;,
\end{equation}
the Onsager angles
\begin{mathletters}
\begin{equation}
	\label{eq:bhat_3}
	\gamma(0) = 2\beta + \ln(\tanh(\beta))
\end{equation}
and
\begin{equation}
	\label{eq:bhat_4}
	\gamma(\pi/L) = \cosh^{-1}(A - \cos(\pi/L))\;,
\end{equation}
\end{mathletters}
where
\begin{equation}
	\label{eq:bhat_5}
        A \equiv \frac{\cosh^2(2\beta)}{\sinh(2\beta)}\;,
\end{equation}
the ``reduced'' mass gap, which is the sum of the mass gaps
for antiperiodic and periodic boundary conditions,
\begin{eqnarray}
	\hat{m}_L (T)& = & [\hat{\xi}_L(T)]^{-1}
		\equiv [\xi_L^a (T)]^{-1} + [\xi_L^{p} (T)]^{-1}
		\nonumber \\
	\label{eq:bhat_6}
	&=& \gamma(\pi/L) - \gamma(0)\;,
\end{eqnarray}
the reduced surface stiffness coefficient
\begin{equation}
	\label{eq:bhat_7}
	\hat{\kappa}_L(T) = \frac{\pi^2} {2L^2 \hat{m}_L}\;,
\end{equation}
and the surface stiffness coefficient in the limit
\( L \rightarrow \infty \)
\begin{equation}
	\label{eq:bhat_8}
	\kappa = \sinh \gamma(0)\;.
\end{equation}
Note that although $L$ was initially allowed to take only
integer values in Eqs.~(\ref{eq:bhat_4})--(\ref{eq:bhat_7}),
we drop this
constraint to find a function
$\hat{\kappa}_L(T)$ which is smooth and differentiable with
respect to $L.$

%==================================================================

Inserting Eqs.~(\ref{eq:bhat_2}--\ref{eq:bhat_8}) into
Eq.~(\ref{eq:bhat_1}), we find
\begin{mathletters}
\begin{eqnarray}
	\label{eq:bhat_9}
	\frac{ d\hat{\kappa}_{L}(T) } {dL}
	& = & \frac{\pi^2} {2L^2 \hat{m}_L}
                \left\{
		\frac{-2}{L} - \frac{1}{\hat{m}_L}
		\frac{d\hat{m}_L}{dL}
		\right\} \nonumber \\
	\label{eq:bhat_10}
	& = & \hat{\kappa}_L
		\left\{
		\frac{-2}{L} - \frac{1}{\hat{m}_L}
		\frac{d\gamma(\pi/L)}{dL}
		\right\} \nonumber \\
	\label{eq:bhat_11}
	& = & \hat{\kappa}_L
		\left\{
		\frac{-2}{L} + \frac{\pi}{ L^2 \hat{m}_L}
		\frac{ d\gamma_(\pi/L) }{d(\pi/L)}
		\right\} \\
	\label{eq:bhat_12}
	& = & -2 \hat{\kappa}_L
		\left\{
		\frac{1}{L}
		- \frac{ \hat{\kappa}_L }{\pi}
		\frac{\sin(\pi/L)}
                     {\sqrt{(A-\cos(\pi/L))^2 - 1}}
		\right\}\;, \nonumber \\
\end{eqnarray}
so that
\mediumtext
\begin{equation}
	\label{eq:bhat_13}
	\hat{b}(L,T)
	  =   \frac{ 2L\hat{\kappa}_L (T) }
		          { \hat{\kappa}_L(T) - \kappa (T) }
		\left\{
		\frac{1}{L}
		- \frac{ \hat{\kappa}_L }{\pi}
		\frac{\sin(\pi/L)}
                     {\sqrt{(A-\cos(\pi/L))^2 - 1}}
		\right\}.
\end{equation}
\end{mathletters}
\narrowtext

%==================================================================

It is convenient to define
\begin{equation}
	\label{eq:bhat_15}
	c \equiv \frac{2}{T_c} = \sinh^{-1}(1)\;.
\end{equation}
A little algebra then yields
the Taylor expansion
\begin{equation}
	\label{eq:bhat_18}
	A \approx  2 + 2c^2t^2\;.
\end{equation}
With this result, and with \(L \! = \! x/t,\)
it is easy to find that
\begin{eqnarray}
	\label{eq:bhat_20}
	\cosh \gamma(\pi/L) & = &
	A - \cos \left( \frac{\pi t}{x} \right) \nonumber \\
	& \approx & (2 + 2c^2t^2) - \left( 1 - \frac{1}{2}
	\left[ \frac{\pi t}{x} \right]^2 \right) \nonumber \\
	\label{eq:bhat_21}
	& = & 1 + \frac{1}{2} g^2(x) t^2 ,
\end{eqnarray}
where for convenience we have defined
\begin{equation}
	\label{eq:bhat_22}
	g(x) = \sqrt{(2c)^2 +
	  \left( \frac{\pi}{x} \right)^2}\;.
\end{equation}
Now taking \(\cosh^{-1} \) of Eq.~(\ref{eq:bhat_21}) gives
\begin{eqnarray}
	\label{eq:bhat_23}
	\gamma (\pi/L) & = & \cosh^{-1} \left(
		A - \cos \left[ \frac{\pi t}{x} \right] \right)
		\nonumber \\
	\label{eq:bhat_24}
	& \approx & g(x)t\;.
\end{eqnarray}
We also expand $\gamma(0)$ in a Taylor series to give
\begin{eqnarray}
	\label{eq:bhat_25}
	\gamma(0) & = & 2\beta + \ln \left( \tanh[\beta] \right)
		\nonumber \\
	\label{eq:bhat_26}
	& \approx & \gamma(0)              \biggl|_{t=0}
		+ \; t \frac{ d\gamma(0) }{dt} \biggr|_{t=0}
		\nonumber \\
	\label{eq:bhat_27}
	& = & 0 + 2ct\;.
\end{eqnarray}

{}From Eq.~(\ref{eq:bhat_8}) this implies
\begin{equation}
	\label{eq:bhat_28}
	\kappa \approx  2ct.
\end{equation}
Using Eqs.~(\ref{eq:bhat_6}), (\ref{eq:bhat_7}),
(\ref{eq:bhat_23}), and (\ref{eq:bhat_26})
provides
\begin{eqnarray}
	\label{eq:bhat_29}
	\hat{\kappa}_L
	& = &  \frac{\pi^2 t^2}
		{2x^2 \{\gamma(\pi/L) - \gamma(0) \} } \nonumber \\
	\label{eq:bhat_30}
	& = & \frac{\pi^2 t}
		{2x^2 \{ g(x) - 2c \} }\;.
\end{eqnarray}
In order to evaluate $\hat{b}(x)$ it is also necessary to
utilize
\begin{eqnarray}
	\label{eq:bhat_31}
	\frac{d \gamma(\pi/L) }{d( \pi/L )}
	& = & \frac{\sin (\pi/L)}
		{\sqrt{\left[ A - \cos(\pi/L) \right]^2 -1}}
		\nonumber \\
	\label{eq:bhat_32}
	& \approx &  \frac{(\pi t / x)}
		{\sqrt{\left[1 + \frac{1}{2} g^2(x)t^2\right]^2 -1}}
		\nonumber \\
	\label{eq:bhat_33}
	& \approx & \frac{\pi}{xg(x)}.
\end{eqnarray}
Then, using Eq.~(\ref{eq:bhat_11})
\begin{eqnarray}
	\label{eq:bhat_34}
	\frac{d \hat{\kappa}_L }{dL}
	& = & \hat{\kappa}_L
		\left\{
		\frac{-2}{L} + \frac{\pi}{ L^2 \hat{m}}
		\frac{ d \gamma(\pi/L) }{d(\pi/L)}
		\right\} \nonumber \\
	\label{eq:bhat_35}
	& \approx & \frac{\pi^2 t}
		{2x^2 [ g(x) - 2c ] }
		\left\{
		\frac{-2t}{x} + \frac{2}{\pi} \hat{\kappa}_L
		\frac{\pi}{xg(x)}
		\right\}.
\end{eqnarray}
Using this and Eq.~(\ref{eq:bhat_30}) we find
\begin{eqnarray}
	\label{eq:bhat_36}
	L \frac{d \hat{\kappa}_L}{dL}
	& \approx & L \frac{\pi^2 t}
		{2x^2 [ g(x) - 2c ] }
		\left\{
		\frac{-2t}{x} + \frac{2 \hat{\kappa}_L }{xg(x)}
		\right\} \nonumber \\
	& \approx & -2 \left\{ \frac{\pi^2 t}
		                    {2x^2 [ g(x) - 2c ] }
		       \right\}
		\nonumber \\
	& &
	\label{eq:bhat_37}  \times
		\left\{ 1 -
		\frac{\pi^2}{2x^2g(x)[g(x) - 2c]}
		\right\}.
\end{eqnarray}
Finally  we can combine Eqs.~(\ref{eq:bhat_1}),
                            (\ref{eq:bhat_28}),
                            (\ref{eq:bhat_30}),
                        and (\ref{eq:bhat_37})
to obtain
\begin{eqnarray}
	\label{eq:bhat_38}
	\hat{b}(x)
	& = & -\frac{L}{\hat{\kappa}_L(T) - \kappa(T)}
	               \frac{d\hat{\kappa}_L(T)} {dL}
		\nonumber \\
	\label{eq:bhat_39}
	& \approx & 2 \left( \frac{\pi^2 t / 2x^2 [ g(x) - 2c ]}
			 { \left\{ \pi^2 t / 2x^2 [g(x) - 2c] \right\}
				- 2ct }
			  \right)
		\nonumber \\
	& &             \times \left( 1 -
			\frac{\pi^2}{2x^2g(x)[g(x) - 2c]}
			\right) \nonumber \\
	\label{eq:bhat_40}
	& = & 2 \left( \frac{1}
			     { 1 - 4cx^2 [g(x) - 2c] / \pi^2 }
			  \right)
		\nonumber \\
	& &             \times \left( 1 -
			\frac{\pi^2}{2x^2g(x)[g(x) - 2c]}
			\right),
\end{eqnarray}
which is correct to \( O(1)\) in $t.$
Since we are interested in $\hat{b}(x)$ in
the limit \( t \rightarrow 0\)
(which for fixed $x$ necessarily means \( L \rightarrow \infty), \)
this is the final answer.

%==================================================================
%\section{Remarks}

It is instructive to determine $\hat{b}(x)$ for
\(x = 0\) and \(x \rightarrow \infty.\)
Since we can write
\begin{equation}
	\label{eq:bIV1}
	xg(x) = \sqrt{(2cx)^2 + \pi^2} \; ,
\end{equation}
we immediately find
\begin{equation}
	\label{eq:bIV2}
	\lim_{x \rightarrow \ 0} xg(x) = \pi.
\end{equation}
Substituting this into Eq.~(\ref{eq:bhat_40}), we find
\begin{equation}
	\label{eq:bIV5}
	\hat{b}(x=0) = 1,
\end{equation}
which agrees with critical scaling \cite{MRS20}.
We also note that
\mediumtext
\begin{equation}
	\label{eq:bIV9}
	\lim_{x \rightarrow \infty} x^2 [g(x) - 2c]
	 =  \lim_{x \rightarrow \infty}
		c \left( \frac{\pi}{2c} \right)^2
		\left[ 1 - \frac{1}{4}
		\left( \frac{\pi}{2cx} \right)^2 \right]\;,
\end{equation}
\narrowtext
so that Eq.~(\ref{eq:bhat_40}) yields
\begin{equation}
	\label{eq:bIV12}
	\lim_{x \rightarrow \infty} \hat{b}(x) = 2.
\end{equation}

%%%%%%%%%%%%%%%%%%%%%%    BIBLIOGRAPHY    %%%%%%%%%%%%%%%%%%%%%

%%%%%%%%%%%%%%%%%%%%%    FIGURE CAPTIONS    %%%%%%%%%%%%%%%%%%%%

\newpage
\narrowtext

%%%%%%%%%%%%%%%%%%%%%%%%%%%%%%%%%%%%%%%%%%%%%%%%%%%%%%%%%%%%%%%%%%%%%%%
\figure{A 3D system illustrating the axes and dimensions of the system,
and the orientation of the interface,
which separates the region of positive magnetization
from the region of negative magnetization. In this figure
boundary conditions are used to tilt the
interface by a fixed angle $\theta$ in the $z$-direction. Although
the boundary conditions discussed in the text do not produce such a
tilt, they can be used to predict the change a tilt would produce
in the surface tension.
\label{fig:3Dsys} }
%%%%%%%%%%%%%%%%%%%%%%%%%%%%%%%%%%%%%%%%%%%%%%%%%%%%%%%%%%%%%%%%%%%%%%%
\figure{The estimates of the surface free energy
in two dimensions.
(a) $\tau_L^a(T),$ from Eq.~(\ref{eq:tau1}), and
(b) $\tau_L^{(+/-)}(T),$ from Eq.~(\ref{eq:tau2}).
The solid curve
is the limit of $\tau_L(T)$ for an infinite lattice \cite{Onsager44}.
The solid vertical line gives the 2D critical temperature.
\label{fig:t2D} }
%%%%%%%%%%%%%%%%%%%%%%%%%%%%%%%%%%%%%%%%%%%%%%%%%%%%%%%%%%%%%%%%%%%%%%%
\figure{The effective scaling exponent $B(tL)$
for $\tau_L^{(+/-)}(T)$
in two dimensions, defined in Eq.~(\ref{eq:tau4}) and evaluated
by numerical differentiation.
\label{fig:B} }
%%%%%%%%%%%%%%%%%%%%%%%%%%%%%%%%%%%%%%%%%%%%%%%%%%%%%%%%%%%%%%%%%%%%%%%
\figure{The estimates of the surface free energy
in three dimensions.
(a) $\tau_{L,M}^a(T),$ from Eq.~(\ref{eq:tau1}), and
(b) $\tau_{L,M}^{(+/-)}(T),$ from Eq.~(\ref{eq:k2d2}).
The dashed curve is a low-temperature expansion \cite{MRS2},
the solid curve comes from series expansion improved with Pad{\'e}
approximants \cite{FishPade}, and small solid
circles (bigger than the error bars)
indicate the results of
Monte Carlo simulations \cite{Pinn92}.
The solid vertical line marks the 3D critical temperature
as determined by a Monte Carlo
Renormalization Group study \cite{MCRGTc}.
\label{fig:t3D} }
%%%%%%%%%%%%%%%%%%%%%%%%%%%%%%%%%%%%%%%%%%%%%%%%%%%%%%%%%%%%%%%%%%%%%%%
\figure{The estimates of the surface stiffness
coefficient in two dimensions.
(a) $\kappa_L^a(T),$ from  Eq.~(\ref{eq:k2d2}), and
(b) $\kappa_{L}^{(+/-)}(T),$ from Eq.~(\ref{eq:k2d8}).
The solid curve is the limit of $\kappa_L(T)$ for an infinite lattice.
The solid vertical line is the 2D critical temperature.
\label{fig:k2D} }
%%%%%%%%%%%%%%%%%%%%%%%%%%%%%%%%%%%%%%%%%%%%%%%%%%%%%%%%%%%%%%%%%%%%%%%
\figure{The effective scaling exponents for the surface stiffness
coefficient in two dimensions.
(a) The scaling exponent $b(tL),$ defined in Eq.~(\ref{eq:k2d4}),
for $\kappa_L^{a}(T).$
(b) The scaling exponent $\hat{b}(tL),$
defined in Eq.~(\ref{eq:bhat_1}), for
$\hat{\kappa}_L(T).$ The solid curve is the limit for large
$L$ referenced in Eq.~(\ref{eq:ap2d5}) and derived in
App.~\ref{app-b}.
\label{fig:b} }
%%%%%%%%%%%%%%%%%%%%%%%%%%%%%%%%%%%%%%%%%%%%%%%%%%%%%%%%%%%%%%%%%%%%%%%
\figure{The estimates of the surface stiffness coefficient
in three dimensions.
(a) $\hat{\kappa}_{L,M}(T),$ from Eq.~(\ref{eq:k3d1}), and
(b) $\kappa_{L,M}^{(+/-)}(T),$ from Eq.~(\ref{eq:k3d1b}).
The solid curve comes from series expansions
improved with Pad{\'e} approximants \cite{FishPade}.
The small solid circles (bigger than the error bars)
are the results of Monte Carlo simulations \cite{Pinn92}.
The vertical dotted line indicates the roughening temperature,
$T_R,$ as determined from Monte Carlo studies \cite{MRS5}.
The vertical solid line is the 3D critical temperature \cite{MCRGTc}.
\label{fig:k3D} }
%%%%%%%%%%%%%%%%%%%%%%%%%%%%%%%%%%%%%%%%%%%%%%%%%%%%%%%%%%%%%%%%%%%%%%%
\figure{The fractional difference between the
transfer-matrix roughening temperature,
$T_R^{\rm TM},$ as determined from
$\hat{\kappa}_{L,M}(T)/k_B T_R = \pi/2,$
and the roughening temperature found from by
Monte Carlo simulations \cite{HasenPhD}, $T_R^{\rm MC}.$
The error bars (shown only for $L \! = \! 3)$ result from
uncertainty in $T_R^{\rm MC}.$
\label{fig:Tr} }
%%%%%%%%%%%%%%%%%%%%%%%%%%%%%%%%%%%%%%%%%%%%%%%%%%%%%%%%%%%%%%%%%%%%%%%
\figure{The estimates for the step free energy using antiperiodic
boundary conditions in the $y$-direction.
Estimates using periodic and free boundary conditions in the
$z$-direction are given by (a) $s^{\rm ap}(T)$ and
(b) $s^{\rm af}(T),$ respectively.
By keeping the temperature constant and
taking 3 different values of $M$ (and any combination
of values of $L),$ we solve Eq.~(\ref{eq:step1})
simultaneously for $s,$
$w,$ and $\mu.$  The solid curve is the 2D surface free energy.
The dashed curve shows the estimate Eq.~(\ref{eq:beta2})
\cite{MRS4,F92PRL}.
\label{fig:s} }
%%%%%%%%%%%%%%%%%%%%%%%%%%%%%%%%%%%%%%%%%%%%%%%%%%%%%%%%%%%%%%%%%%%%%%%
\figure{The estimates for the exponent $w,$ found by solving
Eq.~(\ref{eq:step1}) for $s(T),$ $w,$ and $\mu(T),$ where the
boundary conditions in the $y$-direction are antiperiodic.
It is known that at $T \! = \! 0$ $w \! = \! 1/2$ for
periodic boundary conditions (ap)
in the $z$-direction and that $w \! = \! 0$
for free boundary conditions (af). At low temperatures our
transfer-matrix estimates give that $w \! = \! 0$ for
both sets of boundary conditions, indicating the presence
of substantial corrections to scaling.
\label{fig:w} }
%%%%%%%%%%%%%%%%%%%%%%%%%%%%%%%%%%%%%%%%%%%%%%%%%%%%%%%%%%%%%%%%%%%%%%%
\figure{Finite-size estimates
for the Callan-Symanzik $\beta$ function,
given in  Eq.~(\ref{eq:beta1}).  Results are shown both for
sets of systems with square cross sections and for
extrapolations based on Eq.~(\ref{eq:beta3}).
The solid curve comes from inserting
the estimate of Eq.~(\ref{eq:beta2}) \cite{F92PRL} for $s(T)$
for $T \! \approx \! T_R$
into Eq.~(\ref{eq:step1}). The dashed curve comes from
an approximation of the step free energy by
the 2D surface free energy in Eq.~(\ref{eq:step1}).
\label{fig:beta} }

%%%%%%%%%%%%%%%%%%%%%%%%%%    TABLES    %%%%%%%%%%%%%%%%%%%%%%%%

\newpage
\narrowtext
\begin{table}
  \caption{Boundary conditions discussed in the text.}
  \label{tab:bc}
  \begin{tabular}{lrrr}
       Name & $J_y$ & $h_1$ & $h_L$ \\
       \tableline
       Free         &    0 &   0 &    0 \\
       Periodic     &  $J$ &   0 &    0 \\
       Antiperiodic & $-J$ &   0 &    0 \\
       Plus/Plus    &    0 & $J$ &  $J$ \\
       Plus/Minus   &    0 & $J$ & $-J$ \\
  \end{tabular}
\end{table}

\clearpage
\end{document}